\documentclass[twocolumn,superscriptaddress,showpacs]{revtex4}
\usepackage{graphicx}

\begin{document}

\title{Tailoring wave nonlinearity through spatial composites}

\author{Alessandro Ciattoni} \email{email: alessandro.ciattoni@aquila.infn.it}
\affiliation{Laboratorio Regionale CASTI CNR-INFM, 67010 L'Aquila, Italy}  \affiliation{Dipartimento di Fisica, Universit\`a
dell'Aquila, 67010 L'Aquila, Italy}

\author{Eugenio DelRe}
\affiliation{Dipartimento di Ingegneria Elettrica e dell'Informazione, \\Universit\`a dell'Aquila, 67040 Monteluco di Roio,
L'Aquila, Italy}

\author{Carlo Rizza}
\affiliation{Laboratorio Regionale CASTI CNR-INFM, 67010 L'Aquila, Italy} \affiliation{Dipartimento di Fisica, Universit\`a
dell'Aquila, 67010 L'Aquila, Italy}

\author{Andrea Marini}
\affiliation{Dipartimento di Fisica, Universit\`a dell'Aquila, 67010 L'Aquila, Italy}

\date{\today}

\begin{abstract}
We propose and demonstrate theoretically a method to achieve and design optical nonlinear responses through a light-mediated
spatial hybridization of different standard nonlinearities. The mechanism is based on the fact that optical propagation through a
spatial composite of different nonlinear media is governed by an effective nonlinear response if the spatial scale of the
sequence is much smaller than the light diffraction length.  We apply our general approach to the significant case of
centrosymmetric photorefractive crystal biased by a periodically modulated external voltage to predict strictly bending-free
miniaturized soliton propagation.
\end{abstract}

\maketitle

\noindent

Wave propagation affected by nonlinearity can produce a variety of interesting and useful effects, one of these being the
formation of solitons. In general, these emerge as the product of a specific wave-matter interaction, by which intrinsic wave
features, such as diffraction or dispersion, are countered by the nonlinear response, to produce particle-like dynamics.  In this
respect, solitons are a manifestation of a specific physical context, which can generally be associated with the nonlinear model
that characterizes the wave-matter interaction.

We here focus on the possibility of reversing this perspective, i.e., that of how to produce a prescribed soliton phenomenon, or,
equivalently, of how to devise and tailor an appropriate nonlinearity. A possible solution would be to change the nature of the
material, tailoring its physical and chemical properties, but this in general involves a lengthy material development and does
not provide versatility. We here propose a method to shape, reconfigure, and design a nonlinear response, specialized for the
generation of solitons, based on the macroscopic spatial composition of standard material nonlinearities.  In particular, in
conditions in which this sequencing has a scale that is smaller than the characteristic beam propagation scale (e.g. the
diffraction length), the resulting optical propagation is driven by an effective nonlinearity that, although a product of the
underlying combination, can have different quantitative and qualitative features, i.e., a natural hybridization produced by the
system wave dynamics.

The physical underpinnings are in many respects an elaboration of those that in mechanics are associated with the motion of a
mass particle subject to the combined action of a time-independent force and a rapidly varying external one, a situation first
investigated by Kapitza \cite{Kapitsa1951,Landau}. Here, even though the particle is not able to follow the rapid force
oscillations, yet these produce an additional effective slowly-varying force field. The concept has been successfully extended
and implemented in the general context of nonlinear physics \cite{Chaos}, such as for kink solitons in perturbed sine-Gordon
models, with applications for long Josephson junctions \cite{Kivshar}, guiding-center solitons in perturbed Nonlinear
Schroedinger Equations, with applications in optical pulse propagation in fiber \cite{GuidingCenter}, in quadratic nonlinearities
with applications for spatial quadratic solitons \cite{GuidingCenterQuadratic} and in "nonlinearity management" in both
Bose-Einstein Condensates dynamics and nonlinear optics \cite{mana}. In all these cases, the mechanism is based on a perturbation
of the model propagation equation either through an external modulated source term, or through a modulation of a linear and/or
nonlinear medium property. What we propose here is to bring this Kapitza mechanism even further, by combining and sequencing, at
a fast scale, different media characterized by their own nonlinearities, thus achieving a qualitatively different nonlinear
response (governing the average field) which is not a leading known nonlinearity and corrections due to modulations.

Our proposal can find implementation in a vast variety of systems where macroscopic spatial sequencing is feasible.  For example,
a Kerr layer can be interlaced with a saturable system, such as a liquid-crystal cell, or simply with a Kerr layer with different
nonlinear parameters. An example of a natural setting for this nonlinear designing is found in photorefractive crystals
\cite{Photorefractive}. This is because the nonlinear response is not fixed by the nature of the material itself, but is also the
result of the geometry of the applied external bias voltage. For example, in some conditions the optical nonlinearity can be
switched from a self-focusing saturated response to a self-defocusing one simply by changing the orientation of the field
\cite{Screening}. In distinction to standard nonlinear materials, this malleability allows spatial sequencing in a single
crystal.

Consider the case of the scalar propagation of a paraxial beam through a medium that has the nonlinear response
\begin{equation} \label{deltan}
\delta n(I,I_x,z) = c_0 (I,I_x) +  \sum_{m=1}^{\infty} c_m (I,I_x) \cos \left(\frac{2 \pi m z}{L_v}  \right),
\end{equation}
where $z$ is the propagation direction, $I=|A|^2$ is the optical intensity, $I_x = \partial I / \partial x$ and $L_v$ the
longitudinal scale of variation. Here we have modelled the longitudinal sequence along the $z$-axis of different nonlinear
responses by means of the superposition of a background mean nonlinearity $c_0$ and a periodic nonlinear response whose period is
$L_v$. Note that Eq.(\ref{deltan}) depends on the spatial derivative $I_x$ in order to encompass also slightly nonlocal nonlinear
responses. In the situation where the optical wave is not Bragg-matched with the periodic response \cite{ciatrefl}, the backward
reflected field can be neglected so that the slowly-varying part of the 1D optical field $A(x,z)$ satisfies the parabolic wave
equation $\left[ i \partial / \partial z +(1/2k) \partial^2 / \partial x^2 \right] A = -(k/n_0) \delta n(I,I_x,z) A$ where
$k=2\pi n_0 /\lambda$ (wave-vector carrier), $\lambda$ the wavelength, $n_0$ is the sample background index of refraction. Assume
now that $L_v$ is such that $\lambda \ll L_v \ll L_d$ (the first inequality assuring the paraxial description) so that, retracing
the spirit of the Kapitza mechanical approach \cite{Kapitsa1951,Landau}, it is possible to set $A(x,z)=A_0(x,z)+\delta A(x,z)$,
where $A_0$ is that part of the field that has a longitudinal scale of variation $L_d$, and $\delta A$ is i) longitudinally
rapidly varying, at a scale $L_v$, and ii) it is uniformly in the condition $|\delta A| \ll |A_0|$. This self-consistently
amounts to a decomposition of the field into a slowly varying mean-field component and a rapidly oscillating and small
correction, a direct consequence of condition $L_v \ll L_d$. Substituting this field requirement into the parabolic equation and
exploiting the smallness of $|\delta A|$ as well as the large difference between the two scales $L_v$ and $L_d$ of variation,  it
is possible to derive an equation for $\delta A$ which can be solved and substituted into the equation for $A_0$. The resulting
nonlinear evolution equation for the ''average'' field $A_0$ is
\begin{equation} \label{AverageParabolic}
\left(i\frac{\partial}{\partial z }  + \frac{1}{2k}\frac{\partial^2}{\partial x^2} \right) A_0 = -\frac{k}{n_0} \delta n_{eff}
A_0
\end{equation}
where
\begin{eqnarray} \label{deltaneff}
\delta n_{eff} = && \left\{ c_0 +  \frac{1}{2} \left( \frac{L_v}{\lambda} \right)^2 \sum_{m=1}^{\infty}  \left[\frac{\partial
c_0}{\partial I} I \frac{c_m^2}{m^2} + \right. \right. \nonumber \\ && \left. \left. + \frac{\partial c_0}{\partial I_{x}}
\frac{\partial}{\partial x} \left( I \frac{c_m^2}{m^2} \right) \right] \right\}_{\tiny
\begin{array}{l} I = I_0 \\ I_x = \partial I_0 /
\partial x \end{array} } ,
\end{eqnarray}
\begin{figure}
\includegraphics[width=0.5\textwidth]{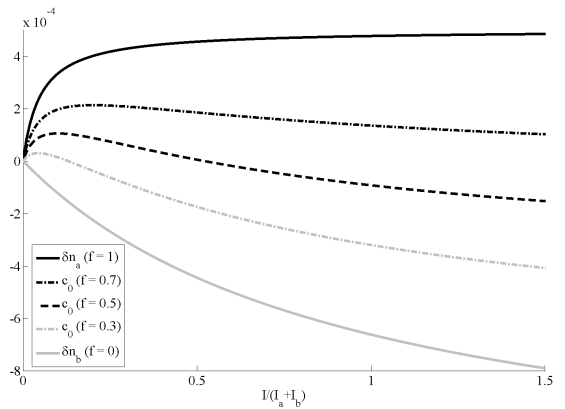}
\caption{Hybridizing a nonlinearity: functional dependence of index change on the intensity, where $\delta n_a = n_a I /(I+I_a)$
and $\delta n_b = n_b I /(I+I_b)$ are two standard saturable nonlinearities, and various effective nonlinear responses $\delta
n_{eff} \simeq c_0$ for different spatial compositions. The result is that different portions of a single wave experience
simultaneously profoundly different self-action, i.e., focusing, defocusing, and even linear (i.e. none). Here we have chosen
$n_a = 5 \cdot 10^{-4}$, $I_a = 0.05 (I_a+I_b)$, $n_b = - 1.3 \cdot 10^{-3}$ and $I_b = 0.95 (I_a+I_b)$.}
\end{figure}
$A_0$ is thus governed by a slowly-varying (i.e., at scales $\gg L_v$) effective nonlinear response $\delta n_{eff}$ which is the
superposition of $c_0$ and a correction consisting in a nontrivial function of $c_0$ and $c_m$ contributions. Therefore, by
appropriately choosing the sequence of longitudinal known nonlinearities, we are in the position to generate effective
nonlinearities which are qualitatively different from their underlying constituents. The validity of the present approach is
guaranteed by the conditions $L_v \ll L_d$ and $(L_v/\lambda) \sum_{m=1}^{\infty} c_m / m \ll 1$ (resulting from the requirement
$|\delta A| \ll |A_0|$) the last inequality being accessible since, even though $L_v \gg \lambda$, the nonlinear coefficients
$c_m$ are generally small. Note that, in Eq.(\ref{AverageParabolic}), the correction to $c_0$ is proportional to $L_v^2$, so, as
we would expect, more rapidly oscillating patterns produces lesser corrective terms to the slowly-varying part of the nonlinear
response.

We are now in a position to fully appreciate the versatility of the idea.  Consider the case of two different saturable nonlinear
media, characterized by the responses $\delta n_a = n_a I /(I+I_a)$ and $\delta n_b = n_b I /(I+I_b)$ ($n_a$ and $n_b$ being the
maximum saturated refractive index changes and $I_a$ and $I_b$ the saturation intensities), sequenced along the $z-$ axis with
the period $L_v$. From Eq.(\ref{deltaneff}) we get $\delta n_{eff} \simeq c_0 = [f \delta n_a + (1-f) \delta n_b]$, where $f$ is
the fraction ($0\leq f\leq 1$) of the period occupied by material $a$. Profiles of $\delta n_{eff}$ are reported in Fig.(1) for a
particular choice of the parameters of the two underlying saturable nonlinearities. Note the non trivial behavior of $\delta
n_{eff}$, being simultaneously self-focusing, self-defocusing, and even linear for different intensities, i.e., for different
parts of the beam: a response that, to the best of our knowledge, does not correspond to any known material optical response, and
whose features (maximum position, width, etc.) can be easily tuned by an appropriate choice of constituents saturable media (i.e.
$n_a$, $I_a$, $n_b$ and $I_b$) and of macroscopic $f$.

\begin{figure*}
\begin{tabular}{ccc}
\includegraphics[width=0.32\textwidth]{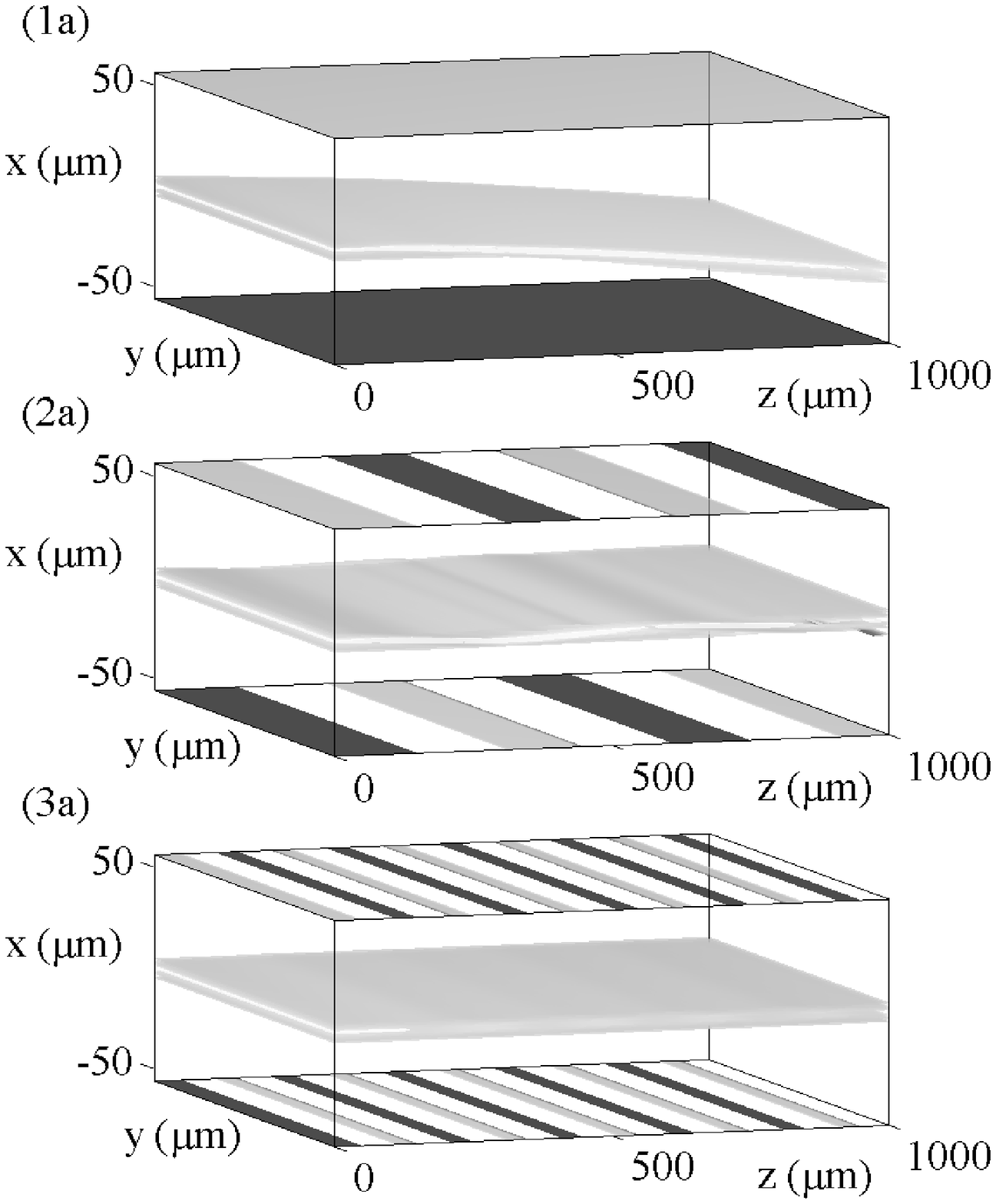} &
\includegraphics[width=0.32\textwidth]{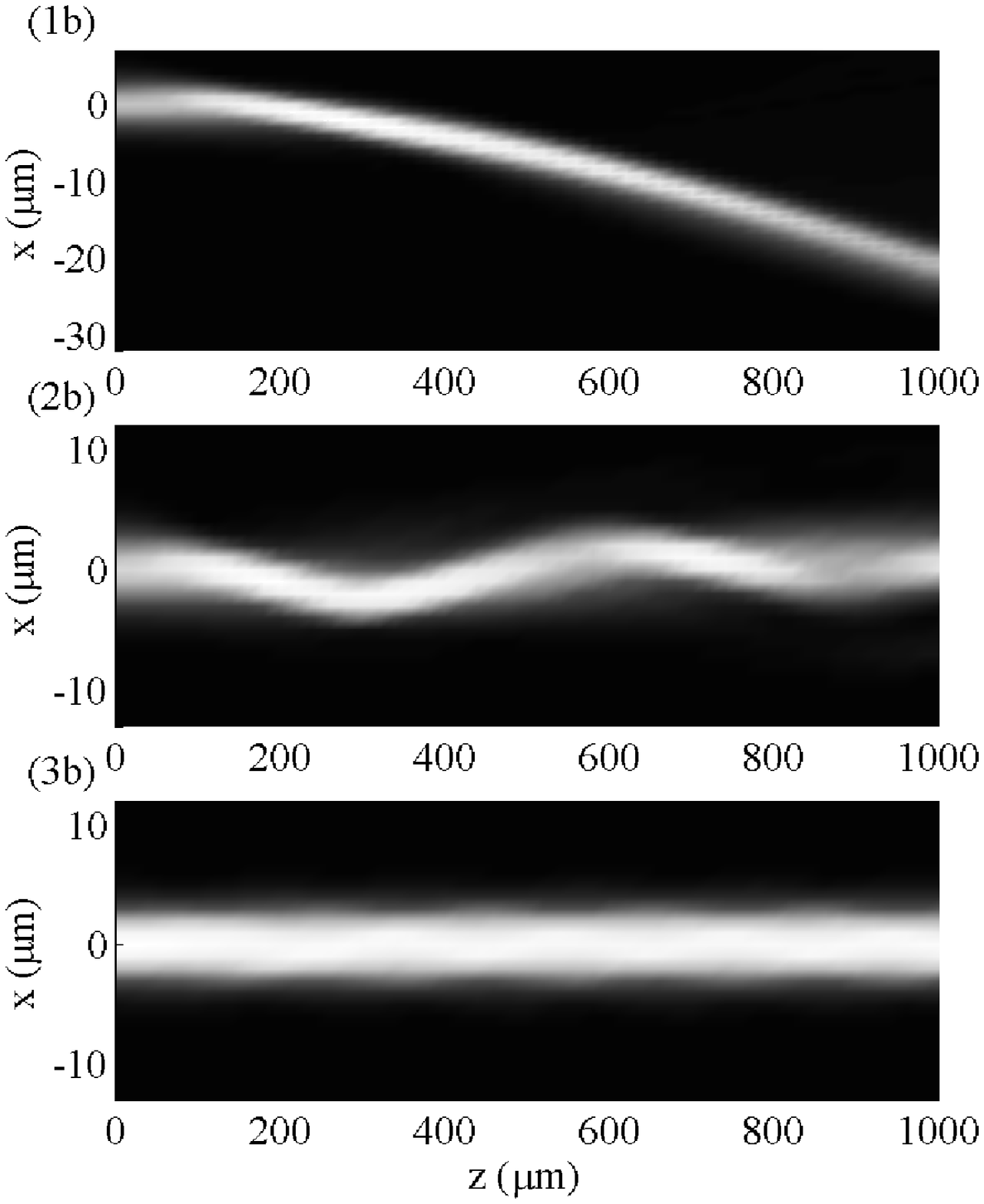} &
\includegraphics[width=0.32\textwidth]{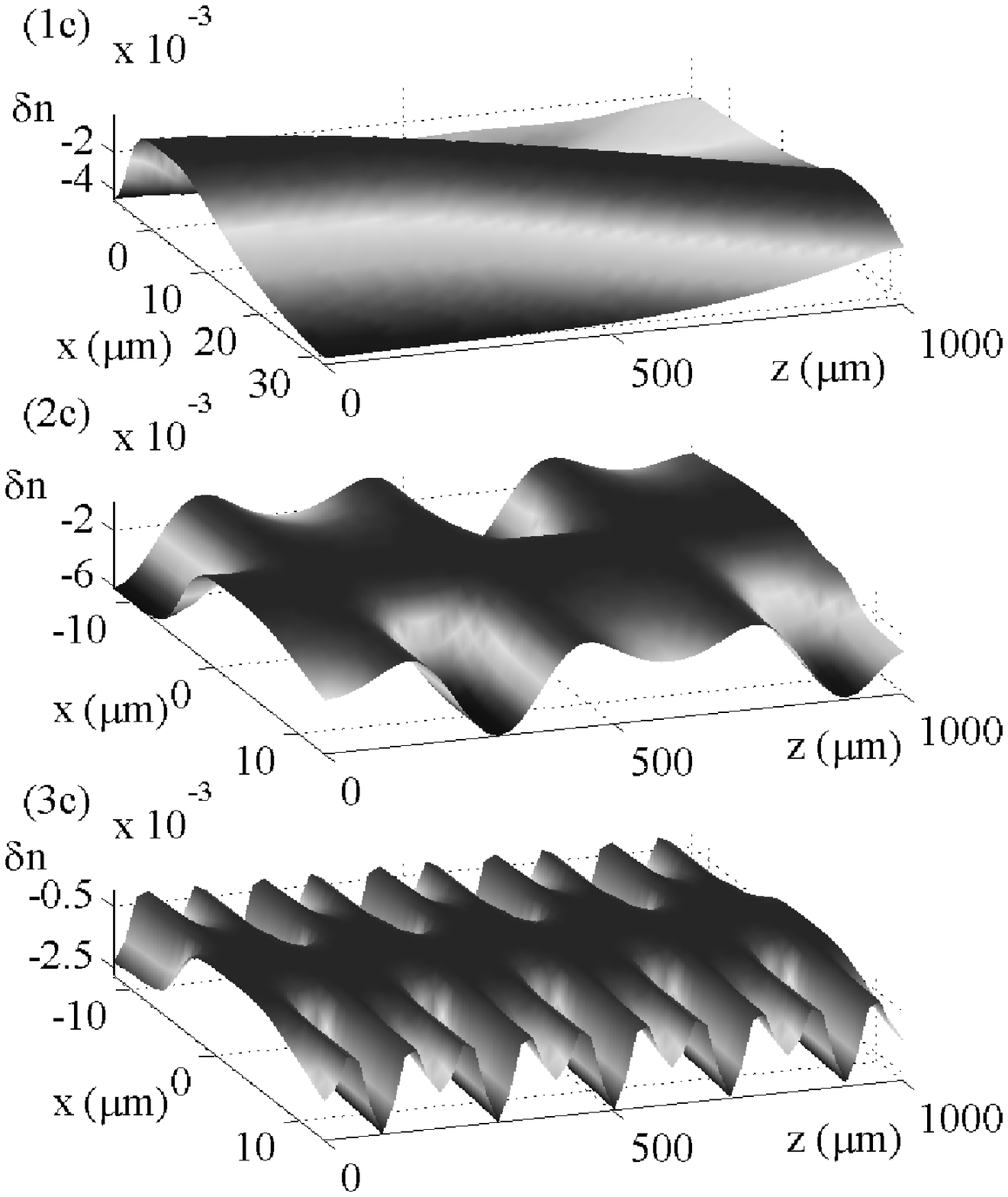}
\end{tabular}
\caption{Devising a bending-free self-trapping nonlinearity in photorefractives. (1a), (2a) and (3a): Crystal layers, electrode
geometries (gray and black stripes) and optical beam (yellow regions) configurations. (1b), (2b) and (3b) : Steady state optical
intensity profiles. (1c), (2c) and (3c) Nonlinear refractive index profiles supporting the corresponding optical propagations.
Note the transition into the Kapitza-like regime in case (3).}
\end{figure*}

In order to pit this method on a pressing issue of some import to recent experiments, we use it to devise an accessible scheme to
achieve photorefractive solitons that do not self-bend \cite{bend1,bend2}. In fact, self-bending impedes such important effects
as miniaturized self-trapping \cite{cha-sat}, i.e., the formation of solitons with micron-sized widths up to the ultimate
paraxial limit, and limits counterpropagating solitons, i.e., head-on collisions or vector soliton composites
\cite{counterpropagating1}. Moreover, bending can have major impact on the interaction of solitons in patterns
\cite{solitons_in_patterns}, because it alters nonlinearly the relative orientation of the beam and the pattern. Consider a
centrosymmetric photorefractive sample as in the three geometries reported in Figs.2-(1a), 2-(2a) and 2-(3a) where the external
bias potential is delivered on the facets $x=L_x$ and $x=-L_x$ by means two standard plate electrodes 2-(1a), or a system of
alternating electrodes along the propagation $z$-axis 2-(2a), 2-(3a). In the (1+1)D configuration at steady state, the light
experiences the nonlinear refractive index change $\delta n= \alpha E_x^2$ where $\alpha = -(1/2) n_0 g \epsilon_0^2 (\epsilon_r
-1)^2$, $g$ is the significant quadratic electro-optic coefficient and $E_x (x,z)$ is the $x-$ component of the electric field
due to the photorefractively stored charges and to the presence of the external bias potential. The associated electrostatic
potential $\Phi(x,z)$ (defined by the relation ${\bf E} = - \nabla \Phi$), in condition where charge saturation can be neglected
\cite{cha-sat}, is governed by the equation $\nabla \cdot [Q \nabla \Phi - \chi \nabla Q] = 0$ where $\chi = K_B T / q$,
$Q=1+I/I_d$, $I(x,z)$ is the optical intensity and $I_d$ is the intensity of a reference background uniform illumination.
Assuming the external bias potential to be of the form $\Phi(\pm L_x,z)=\pm V_0 \cos(\kappa_v z)$, in the condition where the
transverse beam width $\sigma$ (along the $x$- axis) is much smaller than $L_v=2 \pi / \kappa_v$, the equation for $\Phi$ admits
the approximate solution
\begin{equation} \label{potential}
\Phi = \chi \log Q + V_0 \frac{\cosh(\kappa_v x)}{\cosh(\kappa_v L_x)} \left[ \frac{1}{L_x} \int_0^x  \frac{dx'}{Q(x',z)} \right]
\cos(\kappa_v z)
\end{equation}
which, for $\kappa_v=0$ reproduces the well-known one-dimensional photorefractive potential in the standard condition of 2-(1a)
\cite{Screening}. Evaluating the $x$-component of the electric field (retaining only the main contributions and exploiting again
the condition $\sigma << L_v$) we obtain the nonlinear refractive index change
\begin{equation} \label{deltanphot}
\delta n = \alpha \left[ \frac{\chi}{I+I_d} \frac{\partial I}{\partial x} + \frac{\psi}{I + I_d} \cos(\kappa_v z)\right]^2,
\end{equation}
where $\psi = V_0 I_d /(L_x \cosh(\kappa_v L_x))$, so that light dynamics is here governed by a nonlinear response belonging to
the general class of Eq.(\ref{deltan}). Therefore we can apply the above general method to predict that, according to
Eqs.(\ref{AverageParabolic}) and (\ref{deltaneff}), the optical field $A_0$, experiences the effective nonlinear refractive index
change
\begin{equation} \label{deltanNew}
\delta n_{eff} = \frac{\alpha}{2(I+I_d)^2} \left[ \psi^2 + 2 \chi^2 (I_x)^2 \right],
\end{equation}
where we have approximated $\delta n_{eff} = c_0$ (neglecting higher order contributions). Compared to the standard nonlinearity
(i.e., for $\kappa_v=0$), our case has acquired a reflection symmetry ($x \rightarrow -x$) for an even intensity profile
$I(x)=I(-x)$, a symmetry that is intrinsically unavailable for the standard photorefractive response.  The absence of symmetry is
the cause of self-bending for solitons of several microns, but for smaller ones, it actually inhibits their stable formation
\cite{cha-sat,NoteAdded}. It is reasonable to conclude that the we now have the unique possibility of exciting micron-sized
photorefractive solitons, and that these will propagate in a straight line.

To probe the uniform-to-fast-modulation-regime transition, we have numerically integrated the full time-dependent photorefractive
nonlinear optical model \cite{Photorefractive}. In our numerical approach, at each instant of time, we calculate the electric
field distribution induced by the boundary applied voltage and the photoinduced charge solving the $(x,z)$ electro-static Poisson
equation, and the corresponding optical field distribution determined by the electro-optic response through the parabolic
equation \cite{cha-sat}. We have chosen a crystal bulk (layer) of potassium lithium tantalate niobate (KLTN) ($\epsilon _r =3
\cdot 10^4$, $g= 0.13 \: m^4 C^{-2}$, $n_0=2.4$) at room temperature of thickness $2L_x = 2 \times 55 \: \mu m$ and length $L_z =
1000 \: \mu m$ through which an initial Gaussian beam launched at $z=0$ with an half width at half maximum $w_0 = 3.5 \: \mu m$
(with a diffraction length of $\simeq 250 \: \mu m$) and a peak intensity $25 \: I_d$ propagates. In the three electrode
geometries of Figs. 2-(1a), 2-(2a) and 2-(3a), the gray and black stripes are electrodes whose voltages are at $15$ and $-15$
Volts, respectively: in 2-(2a) and 2-(3a) they are alternated along $z$ with periods $600 \: \mu m$ and $200 \: \mu m$. In Figs.
2-(1b), 2-(2b) and 2-(3b) we report the intensity profiles $|A|^2$ of the evaluated optical fields at temporal steady state
whereas in Figs. 2-(1c), 2-(2c) and 2-(3c) we plot the underlying refractive index patterns supporting the corresponding optical
propagations. Configuration (1) corresponds to the standard soliton response where a $3.5$ micron-sized optical beam experiences
a bending deflection $\simeq 25 \: \mu m$ after $\simeq 4$ diffraction lengths. In Figure 2-(1c) we report the corresponding
refractive index, with its curved profile. Note that the beam bends toward the side of negative external potential and, most
importantly, that reversing the applied voltage the resultant beam is the mirror image of the former one since the parabolic
equation with $\delta n$ of Eq.(\ref{deltanphot}) (with $k_v =0$) is left invariant by the transformation $\psi \rightarrow
-\psi$ and $x\rightarrow -x$. When the external applied voltage is slowly modulated along the $z-$ axis (configuration (2)), the
optical beam shows a global wiggling intensity profile. In fact, since the modulation is slow ($L_v > L_d$), the Kapitza
mechanism is ineffective: the situation must be regarded as a cascading of different media, each with an almost uniform
externally bias voltage producing its own asymmetric distortion, characterized by a snake-like response profile (see Fig.2-(2c)).
It is in configuration (3), where the external applied voltage is made to oscillate along the $z-$ axis with a period very close
to the diffraction length, that a wholly different effect arises: the self-trapped beam forms along a straight line.  The absence
of beam deflection is the signature that the Kapitza mechanism, for the nonlinearities, is in action. To underline this, note how
the effective response is not a trivial product of a material response, but is the result of a detailed underlying spatially
oscillating $z$-dependent pattern, as reported in Fig.2-(3c), yet light propagation betrays only a negligible intensity
oscillation (we chose a limiting case with $L_v \sim L_d$). As a final confirmation of the Kapitza picture, i.e. that
configuration (3) is truly governed by the effective nonlinearity, we have integrated both the parabolic equation for $A$ with
the $\delta n$ given by Eq.(\ref{deltanphot}) and the parabolic equation of Eq.(\ref{AverageParabolic}) with $\delta n_{eff}$
given by Eq.(\ref{deltanNew}) with a gaussian boundary (at $z=0$) field profile corresponding to that used in the simulation of
Fig.2-(3a) 2-(3b) and 2-(3c), obtaining, in both cases, good agreement with the results of Fig.2-(3b).

In conclusion, we have devised a general method to achieve an artificial or effective nonlinearity, having light propagate
through a rapidly-varying (but macroscopically accessible) spatial composite of media, each governed by its own nonlinearity. The
fact that the effective nonlinear response can be profoundly different from the underlying nonlinearities of the composite allows
an efficient design of nonlinearity and, in particular, the possibility of a feasible observation of a prescribed soliton
manifestation.  To underline versatility, we have analyzed briefly the composition of two saturable media.  To implement the idea
to overcome a present limitation in soliton investigation, we have considered photorefractive propagation in a modulated external
voltage regime, identifying the conditions for observing micron-sized solitons not suffering the standard deflection arising from
the self-bending effect. This particular result is of fundamental relevance also in applied photonics, since it at once brings us
one step closer to ultimate circuit integration (up to non-paraxial scales) \cite{NonParaxial}, and provides the means to achieve
head-on soliton collisions, the avenue to self-splicing circuits.


\begin{thebibliography} {aa}
\bibitem{NoteAdded} Combination of diffusion and drift yield an off-axis guiding region, since for
$I(x)=I(-x)$, $\delta n$ is the square of the sum of an even ($\alpha \psi/(I+I_d)$) and an odd ($ \alpha \chi I_x /(I+I_d) $)
contribution.
\end{thebibliography}
\end{document}